# Investigating the Requirements for Building a Blockchain-Based Achievement Record System

Bakri Awaji
School of Computing
Newcastle University
Newcastle, UK
b.h.m.awaji2@ncl.ac.uk

Ellis Solaiman
School of Computing
Newcastle University
Newcastle, UK
ellis.solaiman@ncl.ac.uk

Lindsay Marshall
School of Computing
Newcastle University
Newcastle, UK
lindsay.marshall@ncl.ac.uk

## ABSTRACT
A trusted achievement record is a secure system that aims to record and authenticate certificates as well as key learning activities and achievements. This paper intends to gather important information on the thoughts and outlooks of stakeholders on an achievement record system that uses blockchain and smart contract technology. The system would allow stakeholders (for example employers) to validate learning records. Two main aims are investigated. The first is to evaluate the suitability of the idea of building a trusted achievement record for learners in higher education, and to evaluate potential user knowledge of blockchain technology. This is to ensure that a designed system is usable. The second aim includes an interview conducted with a small group of participants to gather information about the challenges individuals have when creating, and reviewing CVs. Overall, 90% of participants agreed that there was a strong need for a trusted achievement record. In addition, 93.64% of respondents stated that they felt it was invaluable to have a system that is usable by all stakeholders. When tackling the second aim it was found that a primary challenge is lack of knowledge of blockchain and its complexity. From the employers' perspective, there is a lack of trust due to inaccuracies when students describe skills and qualifications in their resumes.

## 1. INTRODUCTION
Trusted lifelong learning records that can be used by individuals to present their qualifications and skills to others for various purposes can be extremely valuable. Currently, individuals use traditional methods such as resumes. CVs can be created with the assistance of various online resources, employing a range of structures and styles, or even using social networking websites such as LinkedIn or Facebook. However, the current form of achievement records and CVs has some limitations and challenges that lead to lack of trust [11]. A major problem is poor data continuity. Learning data stays mostly static, even when students move from one institution to a different one. Since institutes have their own independent Learning Record Stores (LRSs), learning data accumulated at earlier facilities cannot be analysed at current, or future institutions. This leads to the cold-start issue under which the existing institution's learning environment lacks enough data to tailor the learning of their students in an efficient way [11].

Employers and other authorities are highly concerned about the validity of academic certificates, due to the institutions providing their certificates no longer being in operation or because they do not maintain accurate records. These situations both lead to difficulty in diploma authenticity validation. As there is a growing number of institutions involved in the global education market, it is increasingly challenging to keep up to date with diploma verification [16].

A further issue is the poor standard of recording and validation of the non-academic achievements of students, which are not stated on any official academic transcripts. These students cannot give a fair and complete representation of their achievements throughout their academic years, such as extra-curricular activities, prizes and employability awards, voluntary work and positions they held in student union clubs and societies [9].

Creating an achievement record is a challenging task, particularly when trying to compile these achievements for a future employer or administrator. It should be noted that numerous individuals put false information on their CVs and claim erroneous achievements on their achievement records. While there are limited studies into resume fraud, it clearly has negative impact [8].

Blockchain [7] is an enabling technology that can play a key role in tackling the problems discussed above. The immutability and security features of blockchain have encouraged researchers to investigate its potential use within a wide array of domains such as banking, cloud computing, IoT, and education. An appealing feature of blockchain technology are smart contracts that can be programmed to automate the process of validating and storing data on the blockchain. A smart contract is an event-condition-action stateful computer program. It can be deployed on top of blockchain to implement a distributed application for multiple parties who do not fully trust one another [17][18]. A more detailed overview of key concepts related to smart contracts and their implementation can be found here [14][10].

### 1.1 Objectives
This study aims to collect valuable data from participants that reflect their opinions and thinking regarding the building of an achievement record system based on blockchain and smart contract technology [4]. Once the data has been gathered, the system requirements can be established, as well as the necessary tools and mechanisms involved. This aim can be fulfilled with a number of objectives:

- To develop user friendly and reliable records for the stakeholders, such as students, employers, and academic staff.
- To gather necessary information from participants of varying educational levels and job roles.
- To offer analytical information on stakeholder outlooks on holding a validated achievement record.
- To establish the difficulties and disparities of achievement record validation.
- To offer analytical information about the average time necessary for the validation of achievement records.
- To present stakeholders with reliable, detailed CVs with all the information they need.
- To offer enough analytical capability to stakeholders in order for them to possess their necessary information.
- To provide analytical information with which achievement records can be developed reliably.
- To define the reasons behind creating a user interface (UI) in the system.
- To pinpoint the data users require in order to show their achievement record.
- To establish how much comprehension of blockchain and smart contract technologies people possess, in order to tailor the user exchange (UX) design of the system.
- To find how difficult the user interface (UI) is to use, in order to tailor a prototype UI.

## 2. METHOD

This study uses a mixed-method approach [12]. It involves the collection of data quantitatively using a questionnaire using a closed set of questions [13]. The questionnaire is created in such a way that numerical data is efficiently collected for further analysis later to produce a generalizable result. The mixed-method approach also involves a qualitative approach through interviews with a number of participants comprised of open questions [6].

The first study was a questionnaire distributed to ≅ 1000 people of varying educational levels and job roles. A response rate of 24.7% (n=247) was achieved, and basic descriptive statistics and specific cross-tabulations were used to analyze the findings. 'SurveyMonkey' was used to design the questionnaire, and section 2.1 presents a detailed analysis of the questionnaire findings [2].

The second study involved interviews with 6 participants responding to a questionnaire to address certain open-ended questions [5]. Section 2.2 presents a detailed analysis of the findings of the interviews. In Section 3, we present a discussion of the two studies conducted in this paper.

### 2.1 Study 1: Questionnaire

The first step was to set a question which was able to evaluate the education level of the participants. Figure 1 shows that the education level is split into eight categories, ranging from 'Less than high school degree' to 'Ph.D. degree or other.' Results show that the respondents have a range of educational levels, meaning that received data is richer as a result of the participants coming from a variety of educational backgrounds. For the current study sample, 38.87% were of bachelor's degree level, 35.22% were of master's degree level (35.22%) and 14.17% were of Ph.D. degree.

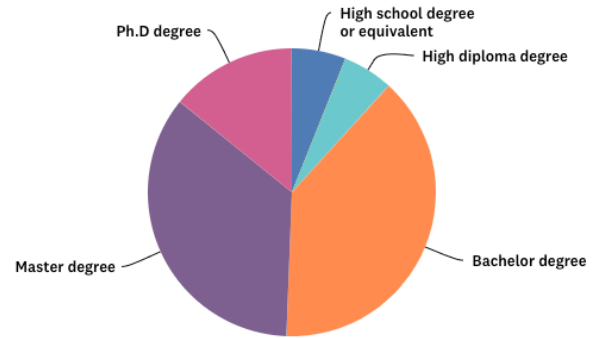

**Figure 1. The respondent's education level.**

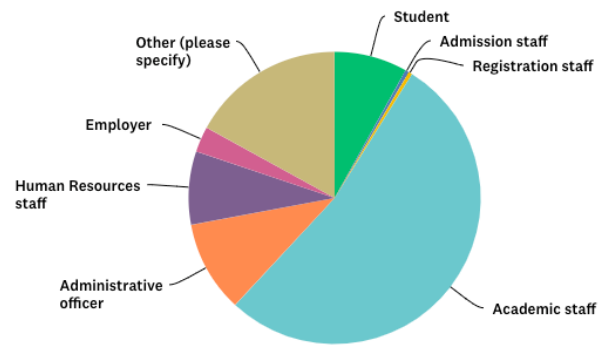

**Figure 2. The occupations of participants.**

Next, the occupations of the study sample were collected, shown in Figure 2. Most of the participants were academic staff (53.043%). Student representation was (8.10%). It is notable that all involved would be beneficiaries of the proposed trusted achievement record.

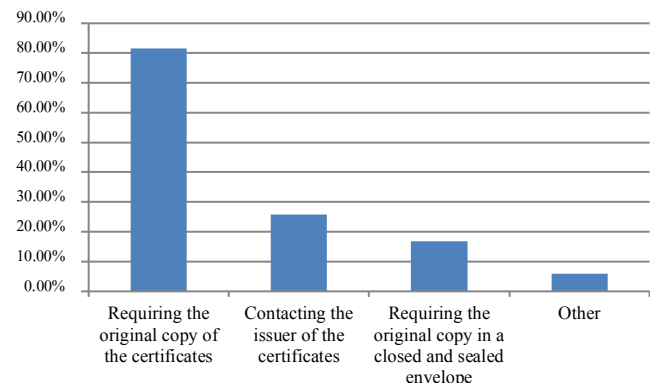

**Figure 3. The methods used to verify data in CVs.**

Validating the information provided in CV is an essential process for stakeholders (for example, employers, administrative staff). In the scope of the current study, the method of verifying data in CVs was investigated. Within the questionnaire, participants were asked the question; "what do you do to check the validity of data in the resume?" 136 answers were received. In Figure 3, it can be seen that 81.62% require the original copy of certificates from the applicants. While 25.74% contact the issuer of the certificates by email, phone, etc. 16.91% require the original copy in a closed and sealed envelope from the issuer. And 5.88% use other methods like digital signatures.

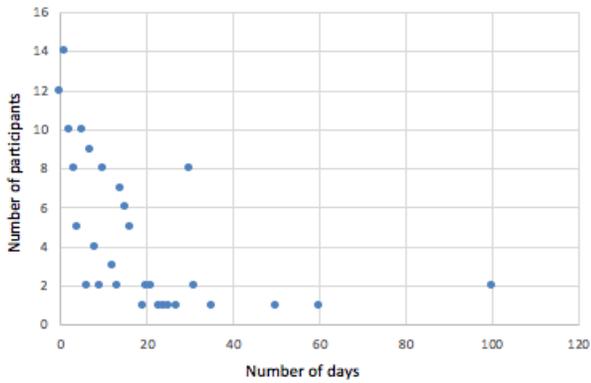

**Figure 4. Time participants take to validate CVs data.**

We also examined the time it takes to validate the correctness of information presented within CVs. Asking the question: "How long does the validation process take?". Figure 4 illustrates the distribution of responses from 131 participants. As the figure shows, on average, record validation takes 12 days.

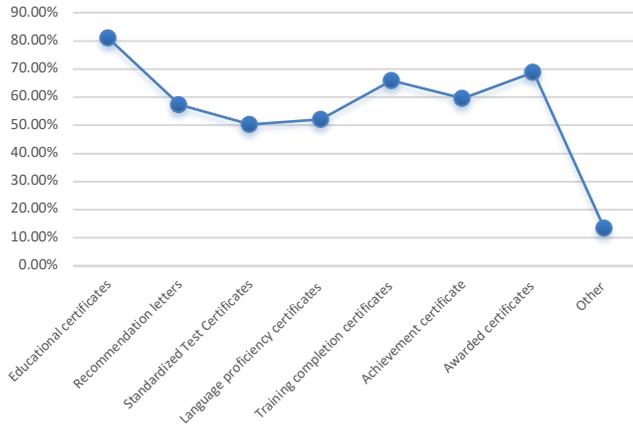

**Figure 5. contents selected to be added to the Achievement Record.**

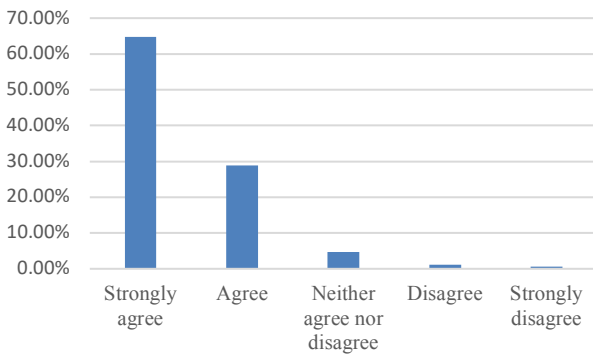

**Figure 6. Trusted achievement record usefulness.**

Next, the participants were asked what the key content of their trusted achievement record would be. Their answers were able to show that the core components of trusted records would be education certificates, letters of recommendation, standardized tests, language ability, training completion, achievement certification and certificates earned. As shown in figure 5.

The participants were then asked to state if they felt that the trusted achieved record is beneficial, and it was shown that 64.74% strongly agreed with this notion, and a total of 93.64% either agreed or strongly agreed. Their responses can be seen in Figure 6.

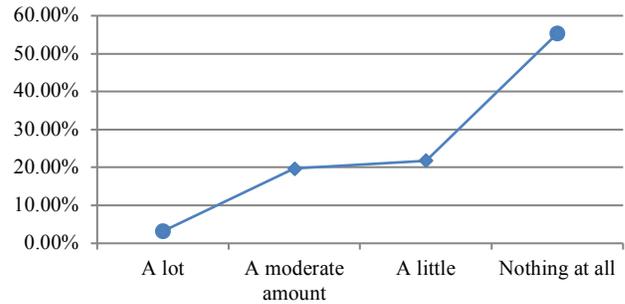

**Figure 7. Knowledge of blockchain technology.**

In Figure 7, it can be seen that the majority of respondents (55.32%) did not have any understanding of blockchain technology, while 80% of participants also said that they have never undertaken a transaction using blockchain technology.

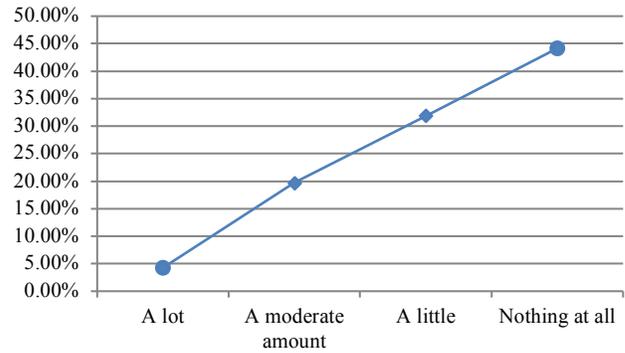

**Figure 8. Participant knowledge about digital wallets.**

The participants were then asked about their awareness of digital wallets. As shown in Figure 8, (44.15%) had no knowledge about digital wallets, and most stakeholders (71.70%) had never used digital wallets, Ethereum wallets or Metamask. On the other hand, (28.30%) of participants had experience of dealing with digital wallets [3].

## 2.2 Study 2: Interview

Interviews were undertaken with specific participants, in order to gather data about the problems they had when developing their CVs, or when trying to use them. Table 1 describes the individuals taking part in the interviews.

**Table 1. interview participants**

| Participant | Occupation | Educational level |
|---|---|---|
| P1 | Student | Bachelor |
| P2 | Registration staff | Bachelor |
| P3 | Admission staff | Bachelor |
| P4 | Academic staff | Ph.D. |
| P5 | Human Resource | Master |
| P6 | Employer | Master |

Their responses were evaluated from active, analytical and critical perspectives, and the data's implications are made clear when trying to understand the experiences shared. Two direct questions were asked, which were; what is the challenge they face when creating their CV, and what are the problems they encounter when reading or reviewing other CVs?

Participant 1 stated that they are often confused about their CV's structure, and that while they can find many templates online, they do not know which would be the best choice. Further issues they had was the amount of time needed to complete their CV and the fact it needed frequent updates. For the second question, they said they did not know. Their responses made it clear that the CV's structure is an area of primary importance to the participants.

P2 responded differently. For the first question they said that they did not find anything difficult, and that it is easy for them to organize their CV with a good structure. To the second question, they said that as a registration employee, they would firstly examine the personal information and address given, to make sure it is updated and that it matches their ID or passport information. This information will be placed in their system and then printed on the certificates. The main observation here is that there is poor accuracy involved sometimes.

P3, responding to question one, said that they can write their CV with no issue. For the second question, they felt that a CV is a necessary document when applying to university, and they have seen a great number of applicant CVs with mistakes, which meant that they requested the original documents. Once again, the observation of these answers is that there is a lack of accuracy.

In response to the first question, P4 said that they had no problem with CVs. Responding to the second question, they felt that people often over-described their current qualifications and skills. Thus, the initial observation for this question is that over-describing occurs.

To the first question, P4 said that they had no issues. For the second question, they said that sometimes people claimed skills on their CV for which they had no proof or certificates, meaning they had difficulty knowing if they were being honest. The observation of this question is that trust issues exist regarding CV information.

P5 firstly stated that writing a CV was easy for them, but it requires time, and this is especially the case for people who have numerous qualifications and abilities, as well as a long work history. To the second question, they said that while they review many CVs on a weekly basis, it is not easy for them to judge the individuals this way. As a result, they ask for an interview, a presentation or a practical test, in order to provide them with a better opportunity to evaluate the person. Also, getting in touch with certificate issuers is a difficult process when trying to verify certifications, and they often delegate this task. Lastly, sometimes CVs include false information, or the author made clear mistakes**.** The initial observations for P5's answers are that there are fairness issues, a need for more time and effort, issues of dishonesty, and additional reactions needed.

For the first question, P6 said that CV creating was not a complicated task for them. In response to the second question, they said that they take their responsibility of choosing employees very seriously, as their success depends on them. For the private sector, skills of applicants are of greater value, so the interviews they conduct are primarily focused on skills. The problem they have is that they need to verify the qualifications on the CV. Also, applicants can over-describe their skills, or even lie, due to the competitiveness involved. The observation for this answer is that there are fairness concerns, extra effort needed, issues with dishonesty, and the problem of over-describing.

## 3. DISCUSSION

### 3.1 Study 1: Questionnaire

Of those who participated in the study, the majority of stakeholders check the validity of CV information before making decisions. An intuitive interpretation is that some people have an immature and less developed approach toward their achievement records. As discussed earlier, CV fraud is pervasive and is predicted to increase even more in the future due to intense competition between job seekers [15]. This is reflected in our results, which demonstrate a lack of trust in the reliability of submitted CVs. When asking stakeholders how they prefer to verify information in CVs, their answers demonstrate a diversity of methods as seen in figure 3, even though applying these validation methods may not be efficient and take time and effort. Improving validation process efficiency in our opinion is an essential future research requirement, which we aim to address through the design and development of a trusted achievement record using blockchain technology [1].

Most respondents (91.71 %) stated that they would like to have a trusted record of all their records of achievement. To determine the context of a trusted achievement record, we asked about essential information and documents that should be stored. A majority of respondents agreed that educational certificates are the essential documents that should be stored. Moreover, participants answered that the record of achievement must contain other important documents (for example; awards, language proficiency docs, test results, etc.). Therefore, diversity of document storage capability is an important factor to consider when developing such a system.

We have demonstrated the level of knowledge among people about technology through an inductive and deductive set of questions. As a result, most of the respondents emphasize that they do not know much about blockchain technology. We asked participants who do have a background in blockchain if they had undertaken any transactions using blockchain. The feedback from most of the respondents reveals that they have no experience of performing a transaction using any digital wallet. Therefore, the ability to manage the complexity of blockchain technology is one of the primary challenges for researchers, and the lack of understanding and knowledge of the technology is a challenge for the end-users as well. Thus, it is essential for future research to address these challenges while developing a blockchain-based system.

### 3.2 Study 2: Interview

We conducted interviews with selected participants to collect more information concerning issues and challenges they are facing in two situations. The first is when they write a CV and the other is when they read and evaluate candidate CVs in order to make recruitment decisions. Thus, we can identify gaps that need to be covered in future research. One of the problems is the diversity in the designs and structures of CVs. This diversity affects quality of the CVs as well as efficiency of analyzing and processing their content. Precision is another problem, especially when people describe their qualifications, tasks, and jobs. Over-describing of skills is a common occurrence, which is due to high competition between job applicants. Consequently, trust in CV content, and fairness when comparing CVs is affected, which is a key problem.

## 4. RECOMMENDATIONS

CV fraud is pervasive and has negative consequences; it hurts organizations/employers, is unfair on qualified applicants with honest CVs, and it can tarnish reputations, increase hiring and training costs to replace terminated fraudsters, propagate unethical cultures, cause poor performance when job-related skills are lacking, or risk legal liabilities related to negligent hiring [11]. The results of the work described in previous sections demonstrates that the majority of our study participants agreed that it is necessary to develop a trusted achievement record capable of addressing the highlighted problems.

In this paper we recommend exploration of the design of a blockchain-based system that provides a trusted achievement record service with consideration to the needs of users, as discussed in section 3. Considering the lack of knowledge in blockchain and its use, as confirmed previously, handling blockchain technology's complex nature should be a key area of attention for researchers. The poor comprehension of this technology is a major hurdle for end-users. Therefore, it is recommended to invest effort in the design of a friendly user interface to facilitate dealing with the complexity and usability of this promising technology.

## 5. CONCLUSION

Credential fraud is becoming more common, and is a major concern that affects educational institutions, employers, and potential employees. Recording and verification of achievements using blockchain technology has plenty of potential offering key advantages, including the potential to offer superior usability and efficiency compared to legacy systems. It is our hope that the findings of this work can serve as a platform to undertake future research towards eliminating credential fraud, and to facilitate the creation of a trusted achievement record that can be easily shared and used by interested stakeholders in a usable, and trusted manner.